\documentstyle[12pt]{article}

\begin{document}
\title{Rescaling of Nuclear Structure Functions}
\author{S.V.Akulinichev\\
{  \it Institute for Nuclear  Research, 60-th October Anniversary}\\
{  \it  Prospect 7a, Moscow 117312, Russia}}

\maketitle
\begin{abstract}
It is shown that
nucleonic structure
functions are $x-$ and $Q^{2}-$rescaled in nuclei.
The $x-$rescaling accounts for nuclear effects in the case of exact scaling,
while the $Q^{2}-$rescaling is
responsible for a corresponding modification of quantum corrections.
This result is  obtained in the leading order for all flavour combinations
and connects the two known models for the EMC-effect.
Electroproduction and gluonic nuclear structure functions
are calculated.

\end{abstract}
\newpage

1. There is still no consensus on the origin
of the  nuclear EMC-effect\cite{EMC}.
In addition to the theoretical importance of this phenomenon, it is now
becoming
practically important to know  nuclear quark and gluon structure
functions (SF) in order to search for novel effects in hadron-nucleus and
in nucleus-nucleus collisions.
Among several models for the EMC effect, we focus on the $Q^{2}$-rescaling
model \cite{JCRR}
and the $x$-rescaling model \cite{AKV}.
In some papers
\cite{SF,Brown} it was concluded that the $x-$rescaling alone
cannot explain the EMC-effect.
An ansatz that both types of rescaling can be combined has been exploited
in Refs.\cite {Ciofi,DT}.
A possible duality of these two models
was discussed in Ref.\cite{Close}, where an interesting
correspondence between the conventional nuclear theory and the QCD
properties of nucleons was conjectured. We show that
there is indeed a deep correspondence between these two types of rescaling
in nuclei but our conclusions differ from the conclusions of
Ref.\cite{Close}. We claim that the $x-$rescaling accounts for the nuclear
modification of nucleonic SF in the tree approximation for
intrinsic partons, while
the $Q^{2}-$rescaling represents a corresponding modification of
quantum corrections. Therefore these two types of rescaling are
self-complementary
and must be taken both into account, but are not equivalent to each other.
We prove the ansatz of Ref.\cite{DT} and reveal the QCD
content of the $(x,Q^{2})-$rescaling. We also extend this model
to flavour singlet combinations and calculate
the gluonic EMC-effect.

2.  We assume that in the region of intermediate and large $x$ the
distances involved are short enough and photons interact
with separated nucleons in nuclei.
We consider only nucleons as nuclear constituents and
discuss below the possibility of other constituents.
The photon-nucleus amplitude
$T_{A}(q,P_{A})$ can be represented in the convolution form
\begin{equation}
T_{A}(q,P_{A}) = \int d^{4}p\:c(P_{A},p)\:f(p)\:T_{N}(q,p),
\end{equation}
where $T_{N}(q,p) = \int d^{4}x\:e^{-iqx}\:<p|T(J(\frac{x}{2})
J(-\frac{x}{2}))|p>$ (here $x$ is the coordinate variable,  but it is
the Bjorken variable elsewhere), $c(P_{A},p)$ are normalization factors
absorbing the flux factors\cite{SF,AS},
$f(p)$ is the four-momentum distribution of nucleons in nuclei.
The Lorentz and flavour indices are omitted for simplicity.
In the covariant Feynman formalism we can write
\begin{equation}
p \equiv P_{A} - P_{A-1} = (m_{N} - E/(A-1),-\vec{P}_{A-1}) = (
m_{N} - \epsilon,\:\vec{p}),
\end{equation}
where $E$ and $\vec{P}_{A-1}$ are
an excitation energy and a three-momentum of the spectator (A-1)-nucleus,
which in contrast to the target nucleus may not be in its ground state,
$\epsilon$ has the meaning of a nucleon separation energy.
{}From $(e,e'p)-$experiments it follows\cite{Frul} that the nucleon
four-momentum
distribution can be approximately described for medium nuclei as
\begin{equation}
|\vec{p}|\leq p_{F}=260 MeV,\:\:\epsilon \approx 35 MeV.
\end{equation}

In the large $(-q^{2})$ limit and  with fixed $x= -q^{2}/2pq$,
the operator product expansion (OPE) can be used to obtain
\begin{equation}
T_{N}(q,p,g(\mu),\mu) = \sum_{i,j} (\frac{2pq}{-q^{2}})^{j}
\:C_{i}^{(j)}(q,g(\mu),\mu)\:
O_{i}^{(j)}(p,g(\mu),\mu),
\end{equation}
where $C_{i}^{(j)}$ and $O^{(j)}_{i}$ are Wilson coefficients and reduced
matrix
elements, respectively, $j$ is the spin of an operator of type $i$,
 $\mu$ is the scale parameter and $g(\mu)$ is the
renormalized coupling constant.
We  now claim that $p$ from
(2) can be substituted in (4). The OPE is used to extract
two factors with two scales:
the target-independent Wilson coefficients
and the reduced matrix elements containing unperturbative
target-dependent effects at the scale $\sim m_{N}$. The condition
for OPE is $(-q^{2})\gg k^{2}$, where $k$ characterises
interactions in a target.
Parton interactions, which fulfill this condition,
are accounted for by the regular part of the amplitudes
(the reduced matrix elements).
With respect to this condition, there is no difference between intra- and
internucleon interactions of partons.
Therefore we  find reasonable to include in the regular part of amplitudes
the interactions of partons from different nucleons as well.
The result of these interactions is  approximately
represented by  off-mass-shell momenta $p$.

The $p-$dependence
of $T_{N}$  appears not only in the factors $(2pq/-q^{2})^{j}$
in (4) (this is accounted for by the $x-$rescaling) but also
in the reduced matrix elements $O_{i}^{(j)}$.
Let us prove that $O_{i}^{(j)}$ are $p$-dependent.
In the region of asymptotic $q$ and $p$, the solution of the renormgroup
(RG) equation for $T_{N}$ is
\begin{equation}
T_{N}(\sigma q,\sigma p, g(\mu),\mu) = T_{N}(q,p,\overline{g}(t,\mu),\mu),
\end{equation}
where $\sigma$ is a rescaling factor,
$t = ln(\sigma)$ and $\overline{g}$ is the running coupling constant.
The solutions of the RG equation for $O_{i}^{(j)}$ and
$C_{i}^{(j)}$ can be written as\cite{Cheng}
\begin{eqnarray}
C_{i}^{(j)}(\sigma p,g(\mu),\mu) &=& C_{i}^{(j)}(p,\overline{g}(t,\mu),\mu)\:
exp(-\int^{t}_{0}dx\:\gamma_{i}^{(j)}(x)),\nonumber\\
O_{i}^{(j)}(\sigma q,g(\mu),\mu) &=& O_{i}^{(j)}(q,\overline{g}(t,\mu),\mu)\:
exp(+\int^{t}_{0}dx\:\gamma_{i}^{(j)}(x)),
\end{eqnarray}
where $\gamma_{i}^{(j)}$ is a corresponding anomalous dimension.
The last equation takes place only if the matrix element
is $p-$dependent. Otherwise
the exponent from the former of the equations (6) will not be canceled
in the expansion (4) for the l.h.s. of (5).
This means that $O_{i}^{(j)}$ must be
$p-$dependent  in order to secure (5).
Since this is true for asymptotic momenta,
this is also true in a general case.
Note that this proof can be done  for arbitrary momenta in the MS-scheme.
{}From the definition of $O_{i}^{(j)}$ and $C_{i}^{(j)}$,
these values depend on momenta squared. By defining
\begin{equation}
\overline{C}_{i}^{(j)}(q^{2},g(\mu),\mu)\equiv C_{i}^{(j)}(q,g(\mu),\mu),\:\:
\overline{O}_{i}^{(j)}(p^{2},g(\mu),\mu)\equiv O_{i}^{(j)}(q,g(\mu),\mu),
\end{equation}
the photon - bound nucleon amplitude can  be rewritten as
\begin{equation}
T_{N}(q,p) = \sum_{i} (\frac{x_{0}}{y})^{-i}
\:\:\overline{C}_{i}(q^{2},g(\mu),\mu)
\:\overline{O}_{i}(p^{2},g(\mu),\mu),\:\:\:\: y=\frac{pq}{m_{N}q_{0}},
\end{equation}
where $x_{0}=-q^{2}/2m_{N}q_{0}$.
The $x-$rescaling due to $y$
takes place always when $(pq)\neq (m_{N}q_{0})$, i.e. even for
free moving nucleons with $p^{2}=m_{N}^{2}$ and in the tree approximation
when $\overline{C}^{(j)}_{i}$ and $\overline{O}^{(j)}_{i}$ are constant.
The $p^{2}$-dependence of $\overline{O}_{i}^{(j)}$
has  nontrivial effects only when $p^{2}\neq m_{N}^{2}$ {\it and}
quantum corrections are included, as we now show.

The $p^{2}$-dependence of $\overline{O}^{(j)}_{i}$
cannot be calculated within the perturbative QCD.
But we can transform this dependence into the calculable
$q^{2}-$dependence of $\overline{C}_{i}^{(j)}$.
For a Lorentz-invariant function of $p^{2}$,
the transformation $p^{2}\rightarrow\sigma^{2}p^{2}$ of external momenta
squared is equivalent to the linear rescaling $p\rightarrow\sigma p$
of those momenta.
Therefore $\overline{O}_{i}^{(j)}$ for bound
nucleons can be found by linearly rescaling nucleonic momenta.
For bound nucleons, the rescaling parameter $\sigma$ is given by
\begin{equation}
\sigma = (\frac{p^{2}}{m_{N}^{2}})^{1/2} = (\frac{(m_{N} - \epsilon)^{2} -
\vec{p}^{\:2}}{m_{N}^{2}})^{1/2}\:<1.
\end{equation}
In the tree approximation the scaling is exact,
the anomalous dimensions vanish and both
$\overline{C}_{i}^{(j)}$ and $\overline{O}_{i}^{(j)}$  become
constant\cite{Cheng}.
This means that the $p^{2}-$dependence of reduced matrix elements takes place
due to quark-gluon loops. If the external nucleonic momenta are linearly
rescaled then loop momenta would also be linearly
rescaled by the same factor $\sigma$. Thus,
the $p^{2}-$dependence of reduced matrix elements can be represented by the
linear rescaling of quark and gluon momenta in bound nucleons.
This conclusion is the starting point in the derivation of the
$Q^{2}-$rescaling in the model of Ref.\cite{JCRR}. The
difference is that in Ref.\cite{JCRR} the
rescaling parameter was not given by (9), but was assumed to be
\begin{equation}
\sigma' = \lambda_{N}/\lambda_{A},
\end{equation}
where $\lambda$ is a confinement size for a corresponding  target.
In contrast to $\sigma$, which can be related to empirical nuclear
values, the parameters $\lambda_{N,A}$  are poorly known and allow to fit
the EMC-effect by the $Q^{2}-$rescaling alone.

Following the conjecture of Ref.\cite{JCRR}, we assume
that the linear rescaling of parton  momenta may be compensated
by the same
rescaling of the normalization scale $\mu$,
\begin{equation}
\overline{O}_{i}^{(j)}(p^{2},g(\mu_{N}),\mu_{N}) \approx
\overline{O}_{i}^{(j)}(m^{2}_{N},g(\mu_{N}/\sigma),\mu_{N}/\sigma).
\end{equation}
Here $\mu_{N}$ is an arbitrary parameter,
interpreted in Ref.\cite{JCRR} as the scale at which the valence-quark
approximation works well for leading twist operators. Our derivation is
not limited by the valence quark approximation. Using the fact that final
results are not sensitive to $\mu_{N}$, if it is of the order of hadronic
mass, we substitute $\mu_{N}=m_{N}$ in numerical calculations.

The derivation of the final equation is now straightforward.
The products $(\overline{C}_{i}^{(j)}\overline{O}_{i}^{(j)})$ are
normalization-independent provided the normalization scale is the same
for both factors.
The $\mu-$dependence of $\overline{C}_{i}^{(j)}$ is known
within the perturbation theory and, in the leading order, is given by
\begin{equation}
\overline{C}_{i}^{(j)}(q^{2},g(\mu),\mu) \sim (\frac{\alpha_{s}(q^{2})}
{\alpha_{s}(\mu^{2})})^{\gamma^{(j)}_{i}/2\beta},
\end{equation}
where $\gamma_{i}^{(j)}$ and $\beta$ are the lowest order gamma and
beta functions. From this equation it follows that
$\overline{C}_{i}^{(j)}(q^{2},g(m_{N}/\sigma),m_{N}/\sigma)=
\overline{C}^{(j)}_{i}(\xi q^{2},g(m_{N}),m_{N})$
and the rescaling parameter $\xi$ is given by
\begin{equation}
\xi =\sigma^{-2\alpha_{s}(m_{N}^{2})/\alpha_{s}(Q^{2})}.
\end{equation}
In the leading order, $\xi$ is
independent of the operator spin and flavour. This means that in this order
all SF moments are rescaled by the same $\xi$ and we can
write the final result in terms of SF,
\begin{equation}
F^{A}(x_{0},Q^{2}) = \int d^{4}p\:f(p)\:(pq/m_{N}q_{0})\:
F^{N}(x_{0}/y, \xi Q^{2}),
\end{equation}
where $F$ can be $F_{2},F_{3}$ or gluon distribution functions (the factor
$(pq/m_{N}q_{0}$) is the reminiscence of the normalization factor in (1)).

We self-consistently rescaled both the
tree-approximation part and the $Q^{2}-$evolution  of SF,
whereas only the latter
is rescaled in Ref.\cite{JCRR}. Consequently, the changes of
scale in these two approaches are different: the average $\sigma$ from (9)
is about 0.94 for the iron nucleus but $\sigma'$ from Ref.\cite{JCRR}
is about 0.87 for the same nucleus, i.e. $(1-\sigma')\approx 2(1-\sigma)$.
Therefore in the approach of Ref.\cite{JCRR} the EMC-effect was fitted
by the $Q^{2}-$rescaling alone, whereas in the present model the
combined $(x,Q^{2})-$rescaling represents the full result.
Note that in Ref.\cite{Close} it was implied that {\it all} parton momenta
(not just valence quark momenta)
would be rescaled by $\sigma'$. In this case bound nucleon momenta, which
are combinations of parton momenta, should also be rescaled by $\sigma'$.
As it follows from (4) and (8),
this leads to an $x-$rescaling as in (14), but with $y=\sigma'$.
However, the numerical results of Ref.\cite{JCRR} leave no room for such
additional $x-$rescaling.

3. We have calculated nuclear SF
using the $Q^{2}-$dependent parametrization of free nucleon SF from
Ref.\cite{Duke} (Set 1) and the nucleon distribution given by (3).
The parameter $\Lambda_{QCD}$ was fixed as in Ref.\cite{Duke}.
The nucleus to nucleon ratios of electroproduction SF are shown in Fig.1.
The $Q^{2}-$rescaling further suppresses nuclear SF at medium and
large $x$.
The description of the
data is reasonable, except for the large-$x$ region. In that region the
contribution of nucleon-nucleon short-range correlations, neglected in this
paper, may improve the agreement with the data\cite{Ciofi}.
The $R$ increasing as $x\rightarrow$1 is automatically accounted for
by the $x-$rescaling in our model, but not in the model of Ref.\cite{JCRR}.
Note that the Fermi motion contribution can disturb the agreement with the
data, obtained in Ref.\cite{JCRR}, even at $x\leq$0.6 (see Ref.\cite{AS}
for the Fermi motion effect).
We conclude that the $(x,Q^{2})-$rescaling can explain the bulk of the
EMC-effect.
In some papers\cite{Gross,Weise}, the $x-$rescaling has been combined
with a modification of nucleonic SF for off-mass-shell
nucleons. That modification gives an additional suppression of nuclear
SF and allows to describe the EMC-effect without
the  $Q^{2}-$rescaling we discussed. There is no contradiction between
the present paper and Refs.\cite{Gross,Weise}: we suggested an alternative
$p^{2}-$dependent modification of SF for bound nucleons, based on the
RG technique. In this approach, the $p^{2}-$dependence of SF takes place only
when the $Q^{2}-$dependence of SF is non-trivial.

The gluonic SF ratios, calculated with only the $x-$rescaling
and with the combined $(x,Q^{2})-$rescaling, are also shown in Fig.1.
The $Q^{2}-$rescaling is very important in this case.
As an illustration,
we applied the calculated quark and gluon nuclear distributions to estimate
the nuclear effectiveness $\alpha$ in the bottomonium production
in proton-nucleus collisions at 800 GeV.
We used the model for this reaction from
Ref.\cite{Akul}.
In this reaction, the dominant contribution is due to gluon-gluon fusion
subprocess.
In Fig.2, $\alpha$ is compared for free nucleon
(dashed line) and nuclear (solid line) SF calculated according to
(14).   The description of the data
is satisfactory except for the region $x\ge 0.25$ ($x_{F}\le 0$). In this
region the comover interaction\cite{Brod}, not taken into account in the
model of Ref.\cite{Akul}, is expected to be significant.
The EMC-effect explains only a small fraction  of the
observed bottomonium suppression but is not negligible.

It is known that the $x-$rescaling
leads to a violation of the momentum sum rule for off-mass-shell nucleons.
This violation can be compensated either by excess pions\cite{ET,AKV} or
by excess gluons\cite{SFL,Akul} in nuclei. The lack of the DY production
enhancement for nuclear targets,
as well as some results for the quarkonium production on nuclei\cite{Akul},
indicates in favour of excess gluons in nuclei.
In Fig.2, we also show our result for $\alpha$ taking into account
the excess gluon contribution, assumed in Ref.\cite{Akul}.
We can say that this contribution does not
contradict the data.
A possible substructure of nuclear binding
forces has been discussed in several papers\cite{Close,SFL,Akul}.
The nuclear rescaling we studied here is not based on
an assumption about the origin of binding forces.

I would like to thank M.Ericson and S.Larin for useful discussions
and the Theory division of CERN for the kind hospitality.

This work was funded by the Russian Foundation of Fundamental Research
(contract No 93-02-14381).

\newpage
{\bf Figure captions.}

Fig.1. The electroproduction cross section ratio for A=56:
the short-dashed line describes
the $x-$rescaling alone, the solid and long-dashed lines describe
the $(x,Q^{2})-$rescaling calculated at $Q^{2}$=10 GeV$^{2}$
and $Q^{2}$=100 Gev$^{2}$, respectively.
The data for electroproduction are from Ref.\cite{EMC}.
The ratio for the gluon distribution functions for A=56:
the dashed-dotted line describes the $x-$rescaling alone and the dotted line
describes the $(x,Q^{2})-$rescaling calculated at $Q^{2}$=100 GeV$^{2}$.

Fig.2 The nuclear effectiveness for  bottomonium production
in proton-nucleus collisions at 800 GeV
calculated with the nuclear structure function from Eq.(14) (solid line)
and with the free nucleon structure functions (long-dashed line).
The former result with the excess gluon contribution included
is shown by the short-dashed line. The data (diamonds for $(1S)-$states
and crosses for $(2S+3S)-$states) are from Ref.\cite{bottom}.


\newpage
\begin{thebibliography}{99}
\bibitem{EMC}J.J.Aubert et al, Phys.Lett. 123B (1983) 275.
\bibitem{JCRR}F.E.Close, R.L.Jaffe, R.G.Roberts and G.G.Ross,
Phys.Rev. D 13 (1985) 1004.
\bibitem{AKV}S.V.Akulinichev, S.A.Kulagin and G.M.Vagradov,
Phys.Lett. 158B (1985) 485.
\bibitem{SF} L.L.Frankfurt and M.I.Strikman, Phys.Lett.
183B (1987) 254.
\bibitem{Brown}G.L.Li, K.F.Liu and G.E.Brown,
Phys.Lett. 213B (1988) 531.
\bibitem{Ciofi}C.Ciofi degli Atti and S.Liuti, Phys.Lett.  225B (1989) 215.
\bibitem{DT}G.V.Dunne and A.W.Thomas, Nucl.Phys. A455 (1986) 701.
\bibitem{Close}F.E.Close, R.G.Roberts and GG.Ross, Phys.Lett. 168B (1986) 400.
\bibitem{AS}S.V.Akulinichev and S.Shlomo, Phys.Lett. 234B (1990) 170.
\bibitem{Frul}S.Frullani and J.Mougey, Adv.Nucl.Phys.
14 (1984)  1.
\bibitem{Cheng}T.-P.Cheng and L.-F.Li, {\it Gauge Theory of Elementary
Particle Physics}, Clarendon Press, Oxford (1984).
\bibitem{Duke}D.W.Duke and J.F.Owens, Phys.Rev. D30 (1984) 49.
\bibitem{Gross}F.Gross and S.Liuti, Phys.Rev. C45 (1992)  1374.
\bibitem{Weise}S.A.Kulagin, G.Piller and W.Weise, Phys.Rev. C50 (1994) 1154.
\bibitem{Akul}S.V.Akulinichev, INR preprint 0887-95(1995)(nucl-th$/$9504032).
\bibitem{bottom}D.M.Alde et al, Phys.Rev.Lett. 66 (1991) 133.
\bibitem{Brod}S.J.Brodsky and P.Hoyer, Phys.Rev.Lett. 63 1989) 1566.
\bibitem{ET}M.Ericson and A.W.Thomas, Phys.Lett. 128B (1983) 112.
\bibitem{SFL}L.L.Frankfurt, M.I.Strikman and S.Liuti, Phys.Rev.Lett.
65 (1990) 1725.
\end{thebibliography}
\end{document}